# Double Sided Watermark Embedding and Detection with Perceptual Analysis

Jidong Zhong and Shangteng Huang

*Abstract*—In our previous work, we introduced a double-sided technique that utilizes but not reject the host interference. Due to its nice property of utilizing but not rejecting the host interference, it has a big advantage over the host interference schemes in that the perceptual analysis can be easily implemented for our scheme to achieve the locally bounded maximum embedding strength. Thus, in this work, we detail how to implement the perceptual analysis in our double-sided schemes since the perceptual analysis is very important for improving the fidelity of watermarked contents. Through the extensive performance comparisons, we can further validate the performance advantage of our double-sided schemes.

Manuscript completed at May 29, 2006. This paper may be taken as a supplement to a paper to be published in IEEE Transactions on Information Forensics and Security. Please refer to [1] for reference. The material presented in this work has also been reviewed by three anonymous reviewers and Prof. Upamanyu Madhow. Thank them for their insightful comments and suggestions.

Jidong Zhong is with the Department of Computer Science and Engineering, Shanghai Jiaotong University, Shanghai 200030, China (e-mail: zhongjidong@sjtu.org).
Shangteng Huang is with the Department of Computer Science and Engineering, Shanghai Jiaotong University, Shanghai 200030, China (e-mail: huang-st@cs.sjtu.edu.cn).

# I. INTRODUCTION

In our previous work [1], we have presented a simple double-sided technique to improve the performance of watermarking systems. It differs from the traditional spread spectrum schemes in that it utilizes the host information at the embedder and detects the watermark by a double-sided decision rule. Furthermore, since our scheme does not reject the host interference at its embedder, it has a nice advantage over the host interference rejection schemes in that the perceptual analysis can be easily incorporated into our scheme. In this work, we continue the discussion in our previous work to investigate the double-sided schemes with perceptual analyses.

Perceptual analyses have been widely employed in watermarking systems to improve the fidelity of watermarked contents. The works [2], [3] exploited the Watson's perceptual model [4] to embed the perceptually shaped signals into the host contents. The Watson's model, initially designed for image compression, includes three major perceptual functions, namely, frequency sensitivity, luminance and contrast masking. Tuned with this model, the image quality is much improved, especially at the smooth regions of the images that are more sensitive to the image manipulations. Since the embedding strength is locally bounded by the Just Noticeable Difference (JND), the maximum possible robustness can be simultaneously achieved. In [5], the authors presented a perceptual model in the DFT domain. The model investigated in [6] exploits the temporal and frequency masking to guarantee that the embedded watermark is inaudible and robust. Similar ideas of using perceptual models to improve both the perceptual quality and the robustness are also reflected in [7]−[13].

In this work, we focus mainly on implementing the widely used Watson's model in the DCT domain for our double-sided schemes. However, it can be similarly extended to other perceptual models. The rest of the work is organized as follows. In Section II, we give some notations that will be used in the later discussions. Section III reviews Watson's perceptual model. In Section IV, we review the previous works that shall serve as a benchmark for performance comparisons. In Section V, we detail how to implement the perceptual analysis in our double-sided scheme. The performance comparisons between our schemes and existing works are reported in Section VI. Finally we conclude the work in the last section.

## II. NOTATIONS

In this work, we denote random variables by italic capital letters and their realizations by italic small letters, such as $X$ and $x$. Random vectors and their realizations are denoted by boldface capital and small letters, respectively, such as **X** and **x**. Other variables are written as italic letters, and vectors as small boldface letters.

Let $\mathbf{X} = \{X_1, X_2, \ldots, X_N\}$ be a collection of $N$ host data, where $X_1, X_2, \ldots, X_N$ are independently and identically distributed (i.i.d.) random variables. In this work, $N$ is supposed to be an even integer. Let $\mathbf{x} = \{x_1, x_2, \ldots, x_N\}$ be a particular realization of the host data. Similarly, the watermarked data and its particular realization are denoted by $\mathbf{S} = \{S_1, S_2, \ldots, S_N\}$ and $\mathbf{s} = \{s_1, s_2, \ldots, s_N\}$ respectively. In the following text, we drop the index to the vector elements when no specific element is concerned. For instance, $X$ may refer to any element $X_i$ in **X**.

Let $\mathbf{w} = \{w_1, w_2, \ldots, w_N\}$ be a bipolar watermark sequence (with $w_i = +1$ or $-1$) and

$$\sum_{i=1}^{N} w_i = 0. \tag{1}$$

The embedder embeds it into the host contents by $\mathbf{s} = \mathbf{x} + g(\mathbf{x}, \mathbf{w})$, where $g$ is an embedding function. The embedded watermark can be detected by testing the hypothesis $H_0$: the test data are not watermarked versus the hypothesis $H_1$: the test data contain the specific watermark **w**. In this work, we characterize the detector's performance by the probability of false alarm ($p_{fa}$) and the probability of miss ($p_m$) which are defined as $p_{fa} = P(\text{say } H_1 | H_0 \text{ is true})$ and $p_m = P(\text{say } H_0 | H_1 \text{ is true})$. Finally, we denote the detector's decision statistic by $L(\mathbf{S})$.

## III. WATSON'S PERCEPTUAL MODEL

Watson [4] originally proposed a perceptual model to design a custom quantization matrix tailored to a particular image. The model consists of three major perceptual functions, i.e., frequency sensitivity, luminance and contrast masking. Let the $(i, j)$th DCT coefficient at block $k$ of the image be denoted by $x(i, j, k)$ with $0 \leq i, j \leq 7$. The mask threshold $m(i, j, k)$ estimates the amounts by which $x(i, j, k)$ may be changed before resulting in any perceptible distortions.

*A. Frequency sensitivity*

Frequency sensitivity describes the human eye's sensitivity to the sine wave gratings at various frequencies. The works [14], [15] provided measurements of thresholds $m(i, j, k)$ for different DCT basis functions. For each frequency $(i, j)$, they measured psychophysically the smallest coefficient that yielded a visible signal. Based on these works, Cox obtained a sensitivity table (see Table 7.2 in Cox [16]). In this work, we also adopt this table for convenience.

*B. Luminance masking*

The frequency sensitivity measures the thresholds for DCT basis functions as a function of the mean luminance of the display. However, the local mean luminance in the image also has a great influence on the DCT thresholds. Watson called this luminance masking and formulated it as

$$m(i,j,k) = m(i,j,k) \cdot [x(0,0,k)/\overline{x}(0,0)]^{a_T}, \tag{2}$$

where $m(i, j, k)$ in the right side of (2) is the frequency sensitivity discussed in the previous subsection, $x(0, 0, k)$ is the mean luminance (or DC coefficient) of the block $k$ of the original image, $\overline{x}(0, 0)$ is the mean luminance of the original image (or the mean of all DC coefficients), and $a_T$ is a constant with a suggested value of 0.649.

*C. Contrast masking*

The threshold for a visual pattern is typically reduced in the presence of other patterns, particularly those of similar spatial frequency and orientation. Watson called this contrast masking and described the threshold for a coefficient as a function of its magnitude, that is,

$$m(i,j,k) = \max\{m(i,j,k),\ |x(i,j,k)|^{w(i,j)} \cdot m(i,j,k)^{1-w(i,j)}\}, \tag{3}$$

where $m(i, j, k)$ in the right side of (3) is the threshold for luminance masking obtained in the previous subsection, and $w(i, j)$ is an exponent that lies between 0 and 1 with a typical empirical value of 0.7. Since the threshold $m(i, j, k)$ roughly scales proportionally to the host coefficient $x(i, j, k)$, the multiplicative embedding rule can thus automatically achieve a nice perceptual quality.

IV. PREVIOUS SINGLE-SIDED WORKS

Podilchuk and Zeng [2] are among the first to take the perceptual quality into account. With the perceptual analysis implemented, the embedding rule for the Additive Spread Spectrum (ASS) scheme can be formulated as

$$s_i = x_i + am_i w_i, \quad (4)$$

where $i = 1, 2, \ldots, N$, $m_i$ is the mask threshold determined in Section III, and $a$ is the embedding strength. The early methods detect the embedded watermark by a simple linear correlator given by

$$L(\mathbf{S}) = \frac{1}{N} \sum_{i=1}^{N} S_i w_i. \quad (5)$$

Though the correlator is not optimum for the host signals that do not follow a Gaussian distribution, it is still widely used in the literature.

Later, Hernandez [12] designed an optimum detector (in the Neyman-Pearson sense) by assuming that the host DCT coefficients follow a Generalized Gaussian Distribution (GGD) whose probability density function (pdf) is given by

$$f_X(x) = A \cdot \exp(-|\beta x|^c), \quad (6)$$

where $\beta = \sqrt{\Gamma(3/c)/\Gamma(1/c)}/\sigma_X$, $A = \beta c / [2\Gamma(1/c)]$, $\sigma_X$ is the standard deviation of the host signal $X$, and $\Gamma(\cdot)$ is the Gamma function defined as $\Gamma(x) = \int_0^\infty t^{x-1} e^{-t} dt$. In (6), $\beta$ is the scale parameter and $c$ the shape parameter estimated from the host image [19]. Gaussian and Laplacian distributions are just special cases of (6) with $c = 2.0$ and $1.0$, respectively. Based on this assumption, an optimum detector [12] can be formulated as

$$L(\mathbf{S}) = \frac{1}{N} \sum_{i=1}^{N} |S_i|^c - |S_i - am_i w_i|^c. \quad (7)$$

In [13], the authors described the host signals by a Cauchy distribution whose pdf is given by

$$f_X(x) = \frac{1}{\pi} \frac{\gamma}{\gamma^2 + (x - \delta)^2} \quad (8)$$

where $\gamma$ is the scale parameter and $\delta$ is the location parameter. For a symmetric pdf, $\delta = 0$. In this work, we focus primarily on symmetrical data. Thus, the above pdf is simplified as

$$f_X(x) = \frac{1}{\pi} \frac{\gamma}{\gamma^2 + x^2}. \tag{9}$$

The scale parameter can be estimated by [17]. In [13], the authors gave a Cauchy detector (LMP detector) with a decision statistic

$$L(\mathbf{S}) = \sum_{i=1}^{N} \frac{2(S_i - \delta)w_i}{\gamma^2 + (S_i - \delta)^2}. \tag{10}$$

Since $\delta = 0$ in this work, the above statistic can be rewritten as

$$L(\mathbf{S}) = \frac{1}{N} \sum_{i=1}^{N} \frac{S_i w_i}{\gamma^2 + S_i^2}. \tag{11}$$

It is also important to note that only frequency sensitivity and luminance masking are considered in [12], [13]. Though the authors call their detectors optimal, they are not strictly so since $m_i$ is taken as a fixed value in their derivation of the optimum detectors. In real scenarios, $m_i$ is image-dependent and not fixed. However, their detectors do achieve a performance improvement over linear correlators if only frequency sensitivity and luminance masking are considered. This is because $m_i$ does not vary much and thus can be roughly taken as fixed. Nevertheless, if the contrasting masking were also implemented, $m_i$ would be dependent on $x_i$ (See (3)) and their detector would not be optimal at this time. Instead, we will see in the future experiments that their detectors yield a very poor performance in such a case.

## V. DOUBLE-SIDED SCHEMES

### A. Double-sided Additive Spread Spectrum Schemes (DS-ASS)

Incorporating the perceptual analysis into DS-ASS, we obtain an embedding rule [1]

$$s_i = x_i + am_i w_i, \text{ if } \bar{x} > 0;\ s_i = x_i - am_i w_i, \text{ if } \bar{x} \leq 0, \tag{12}$$

where $i = 1, 2, \ldots, N$ and $\bar{x}$ is the projected $\mathbf{x}$ on $\mathbf{w}$ defined as

$$\bar{x} = \frac{1}{N} \sum_{i=1}^{N} x_i w_i. \tag{13}$$

Please refer to [1] for further information. The linear correlator with the double-sided decision rule [1] can be employed for its watermark detection, that is,

$$|L(\mathbf{S})| > \psi \Rightarrow H_1; \text{otherwise} \Rightarrow H_0 \tag{14}$$

where $L(\mathbf{S})$ is given by (5) and $\psi$ is the decision threshold.

### B. Double-sided Cauchy scheme

It is difficult to propose a double-sided scheme for the optimum detector designed by Hernandez [12] since their detector requires the knowledge of embedding strength. However, it is possible to design a Double-Sided Cauchy (DS-Cauchy) scheme for Briassouli's scheme [13] with an embedding rule

$$s_i = x_i + am_i w_i, \text{ if } \bar{x} > 0; \ s_i = x_i - am_i w_i, \text{ if } \bar{x} \leq 0, \tag{15}$$

where $i = 1, 2, \ldots, N$ and

$$\bar{x} = \frac{1}{N} \sum_{i=1}^{N} \frac{x_i w_i}{\gamma^2 + x_i^2}. \tag{16}$$

We employ the double-sided decision rule (14) for watermark detection, however with $L(\mathbf{S})$ replaced by (11). To see how the embedding rule (15) works, substituting it into (11), we obtain

$$L(\mathbf{s}|H_1) = \frac{1}{N} \sum_{i=1}^{N} \frac{(x_i \pm am_i w_i) w_i}{\gamma^2 + (x_i \pm am_i w_i)^2} = \frac{1}{N} \sum_{i=1}^{N} \frac{x_i w_i}{\gamma^2 + (x_i \pm am_i w_i)^2} \pm \frac{1}{N} \sum_{i=1}^{N} \frac{am_i}{\gamma^2 + (x_i \pm am_i w_i)^2} \tag{17}$$

Since $a \cdot m_i$ is usually much smaller than $x_i$, thus

$$L(\mathbf{s}|H_1) \approx \frac{1}{N} \sum_{i=1}^{N} \frac{x_i w_i}{\gamma^2 + x_i^2} \pm \frac{1}{N} \sum_{i=1}^{N} \frac{am_i}{\gamma^2 + (x_i \pm am_i w_i)^2} = \bar{x} \pm \frac{1}{N} \sum_{i=1}^{N} \frac{am_i}{\gamma^2 + (x_i \pm am_i w_i)^2} \tag{18}$$

Therefore, we see from (15) and (18) that the second term of (18) is of the same sign with $\bar{x}$, which is the essence of double-sided schemes.

## VI. Discussions

### A. Performance in the sense of Mean Squared Errors (MSE) metric

In real scenarios, $m_i$ depends on the host signal $x_i$. However, to simplify the discussion, we assume that $m_i$ is fixed in this subsection. Thus, the performance of DS-ASS can be described as

$$p_m = \begin{cases} 1 - 2Q[Q^{-1}(p_{fa}/2) - k\sqrt{N}/\sigma_X], & \text{if } \psi > k \\ 0.0, & \text{if } \psi \leq k \end{cases}$$

where

$$k = \frac{1}{N}\sum_{i=1}^{N} a \cdot m_i.$$

Please refer to the previous work [1] for the detailed derivation of the above equation. In the sense of MSE metric, we may assume that the embedding distortion $D_w$ is

$$D_w = \frac{1}{N}\sum_{i=1}^{N} a^2 m_i^2.$$

By Cauchy's inequality, we have

$$k^2 = \left(\frac{1}{N}\sum_{i=1}^{N} am_i\right)^2 \leq \frac{1}{N}\sum_{i=1}^{N} a^2 m_i^2 = D_w,$$

where the equality holds at $m_1 = m_2 = \ldots = m_N$. That is, the best performance of DS-ASS is achieved at $m_1 = m_2 = \ldots = m_N$. Hence, the perceptual shaping reduces the performance of DS-ASS at the same distortion level (in the sense of MSE metric). It also implies that performance analyses without taking perceptual quality into account may be quite misleading. Of course, the above rationale holds similarly for ASS schemes.

### B. Disadvantages of host interference rejection schemes

The host interference rejection schemes have been adapted to watermark detection problems and proven to be superior in performance [20], [21]. However, they haven't taken the perceptual analysis into account. Indeed, it is difficult to implement the perceptual analysis for them. In this subsection, we conjecture a possible way to implement the perceptual analysis for Spread Transform Dither Quantization (STDM) scheme [21] and then discuss its disadvantages.

Suppose that the maximum allowable embedding strength for the host data is $a_{\max}$. The perceptual analysis for our schemes can be simply implemented as $s_i = x_i \pm a_{\max} m_i w_i$ to achieve the maximum possible embedding strength. With the help of the spread transform technique, we can also obtain the embedding rule for STDM with perceptual analysis. Projecting both sides of (4) on **w**, we have

$$\bar{s} = \bar{x} + a \frac{1}{N} \sum_{i=1}^{N} m_i \tag{19}$$

where both $\bar{x}$ and $\bar{s}$ are defined as (13). Let $\bar{s} = q_\Lambda(\bar{x})$, where $q_\Lambda(\cdot)$ is a Euclidean scalar quantizer of step size $\Delta$ whose centroids are defined by the points in the shifted lattice $\Lambda \triangleq \Delta \mathbb{Z} + \Delta/2$ (the $\Delta/2$ is chosen by symmetry reasons) [Please see [21] for details]. Thus we immediately have

$$a = \frac{q_\Lambda(\bar{x}) - \bar{x}}{\left(\sum_{i=1}^{N} m_i\right)/N}. \tag{20}$$

Substituting (20) into (4) leads to an embedding rule for STDM scheme with perceptual analysis. A recent paper [18] proposed to implement the perceptual analysis in STDM scheme for watermark decoding problems. The underlying idea is much similar to our conjecture. Now, we examine the disadvantages of the above embedding rule. In the above equation, $a$ should be smaller than $a_{max}$ to keep the distortion imperceptible. However, if it is smaller than $a_{max}$, it leaves a larger perceptual allowance for the attacker. One may argue that we can select an appropriate step size to make the embedding strength $a = a_{max}$. However, it is not workable since in such a case the detector has no chance of knowing what the step size is used. In fact, the major inherent problem with the host interference rejection schemes is that it is hard to control the embedding strength since it depends on $\bar{x}$ over which we have no control.

## VII. EXPERIMENTAL RESULTS

In this section, we make performance comparisons between the above single-sided and double-sided schemes. The host data come from the fifth coefficients (in Zigzag order) of Lena, Peppers, Boat and Baboon. For Lena, $c = 0.69$ and $\sigma_X = 19.74$; for Peppers, $c = 1.03$ and $\sigma_X = 16.04$; for Boat, $c = 0.73$ and $\sigma_X = 24.32$; for Baboon, $c = 1.11$ and $\sigma_X = 32.12$. The estimated scale parameter $\gamma$ for Lena is 6.69, for Peppers is 7.35, for Boat is 8.65 and for Baboon is 15.59. As also explained in the previous work [1], we permute randomly the host DCT coefficients to obtain different **x** for tests. In the following experiments, the empirical results are obtained over 1,000,000 such permutations. Finally, in the legends of the figures,

"ASS-COR" stands for the linear correlator (5) for ASS, "Hernandez" for Hernandez's detector (7) for ASS, "Briassouli" for Briassouli's detector (11) for ASS, "DS-ASS" for the scheme (12) and "DS-Cauchy" for the scheme (15).

*A. Perceptual analysis with frequency sensitivity and luminance masking*

As done in [12], [13], only frequency sensitivity and luminance masking are considered in this subsection. The experimental results are displayed in Fig. 1 to Fig. 3. In these figures, we observe that DS-Cauchy does achieve a better performance than its single-sided counterpart. In accordance with the conclusions drawn in the previous work [1], DS-ASS also improves ASS-COR. However, it is really hard to decide whether DS-ASS is better than DS-Cauchy. DS-Cauchy can achieve a better performance at the low false alarm probabilities since it is based on a locally most powerful detector. Nevertheless, as explained in the previous work [1], DS-ASS is much simpler and entails no parameter estimation. Finally, we observe from these results that Hernandez's and Briassouli's schemes enjoy a better performance than ASS-COR. This is because $m_i$ does not vary much and can be roughly taken as fixed.

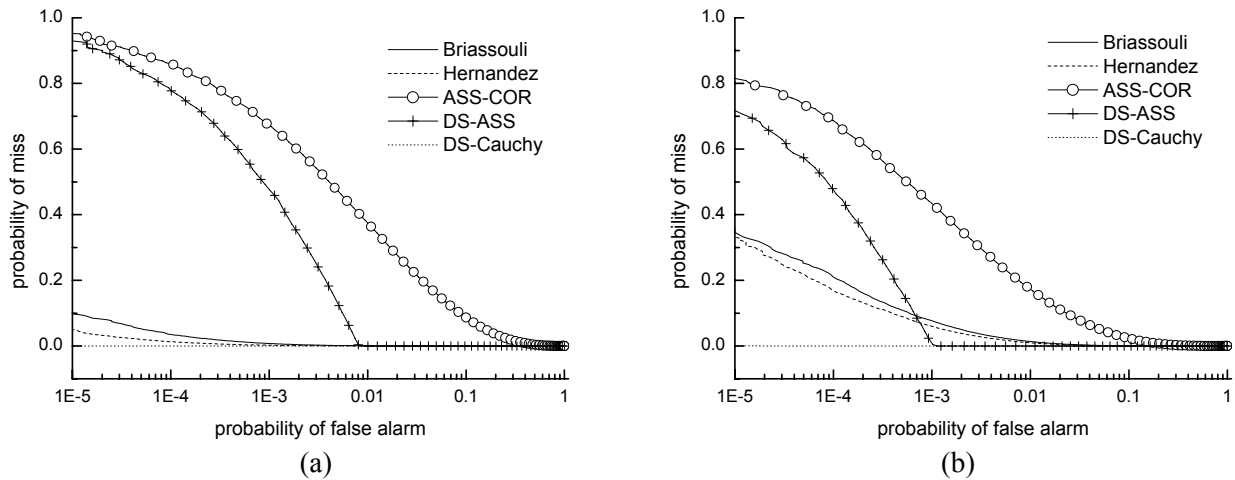

(a)  (b)

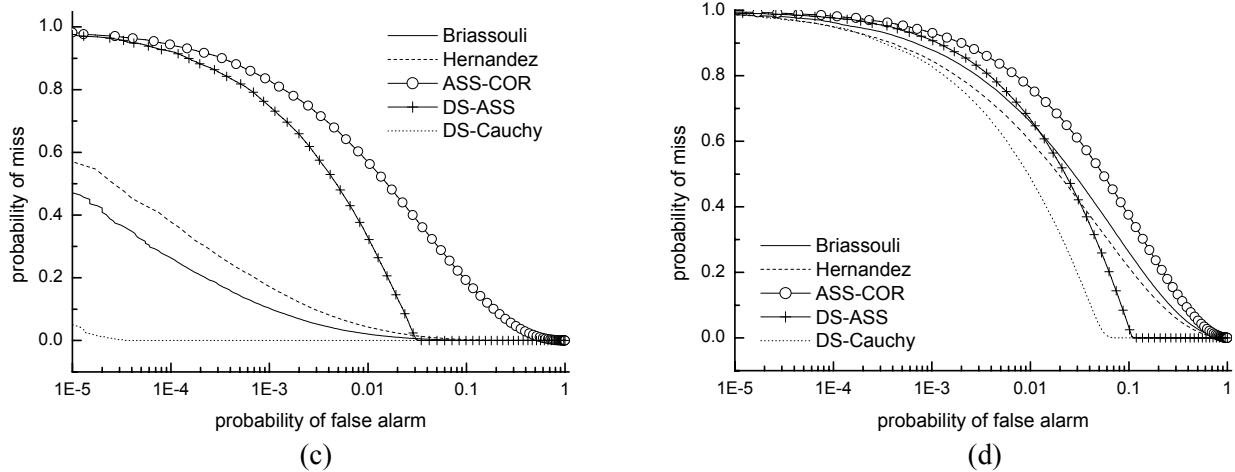

Fig. 1. Performance comparisons under no attack (with $a = 1.0$, $N = 2000$). (a) Lena. (b) Peppers. (c) Boat. (d) Baboon.

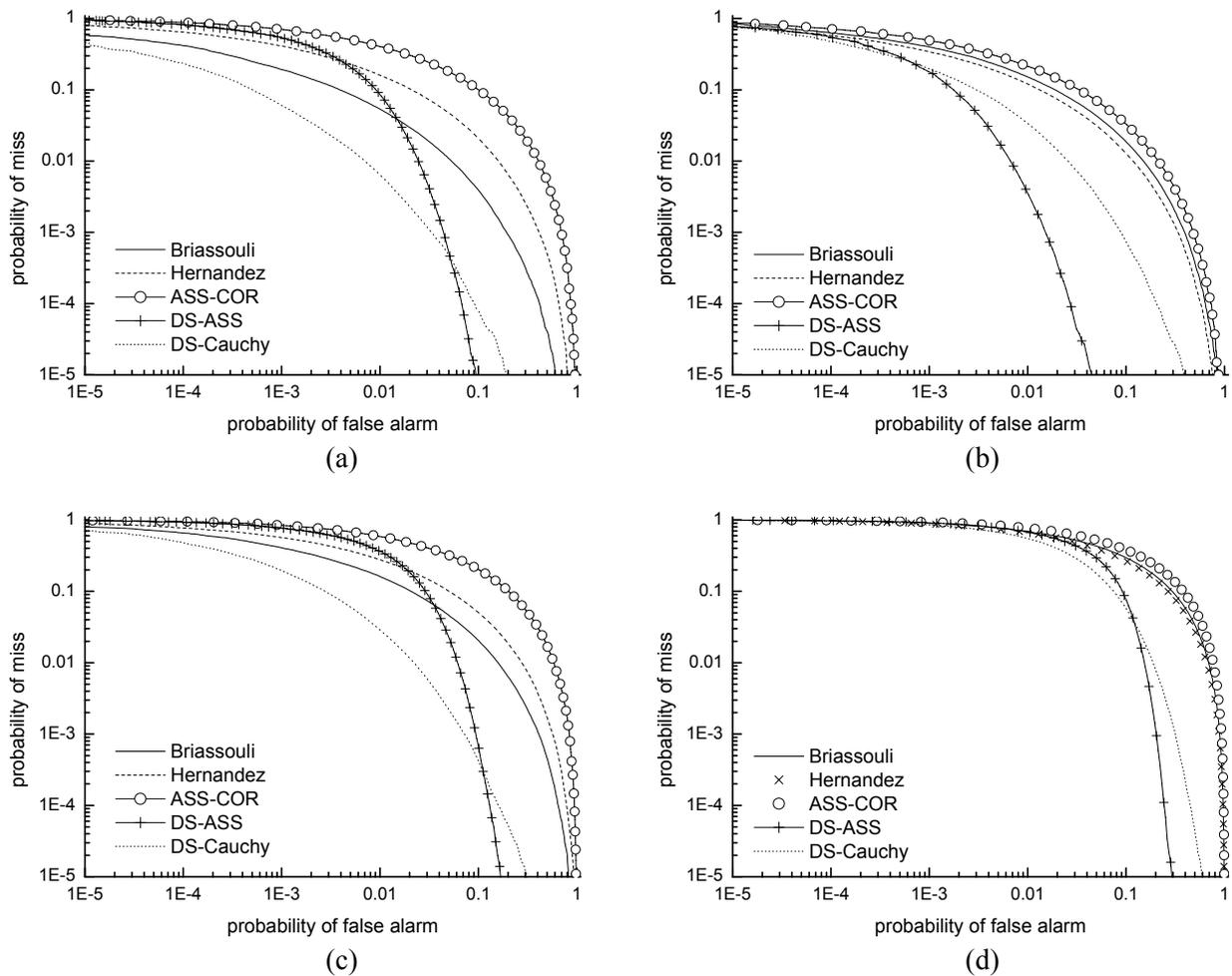

Fig. 2. Performance comparisons under zero-mean Gaussian noise attacks $\mathcal{N}(0, 25)$ (with $a = 1.0$, $N = 2000$). (a) Lena. (b) Peppers. (c) Boat. (d) Baboon.

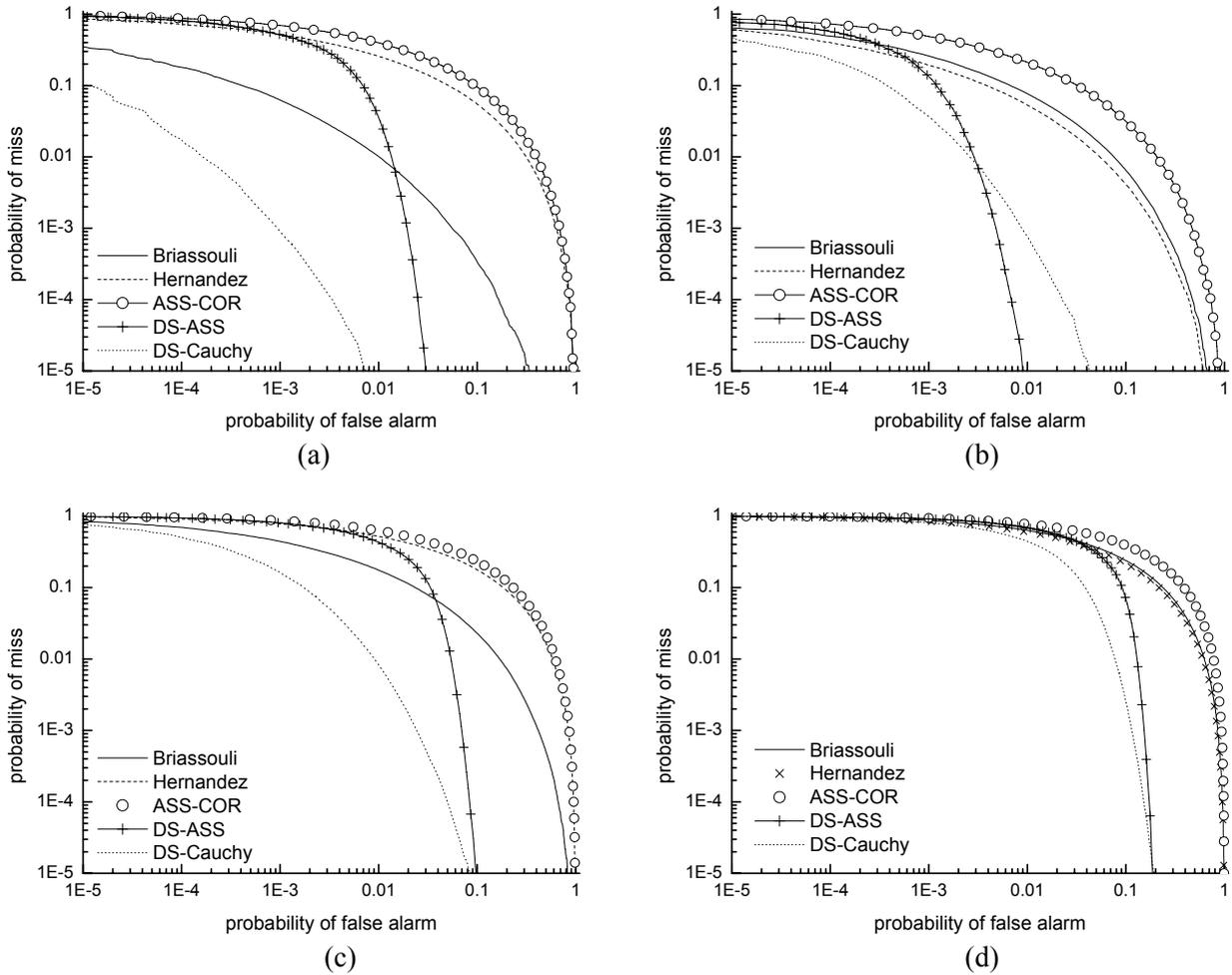

Fig. 3. Performance comparisons under JPEG (QF = 50) attacks (with $a = 1.0$, $N = 2000$). (a) Lena. (b) Peppers. (c) Boat. (d) Baboon.

*B. Perceptual analysis with frequency sensitivity, luminance and contrast masking*

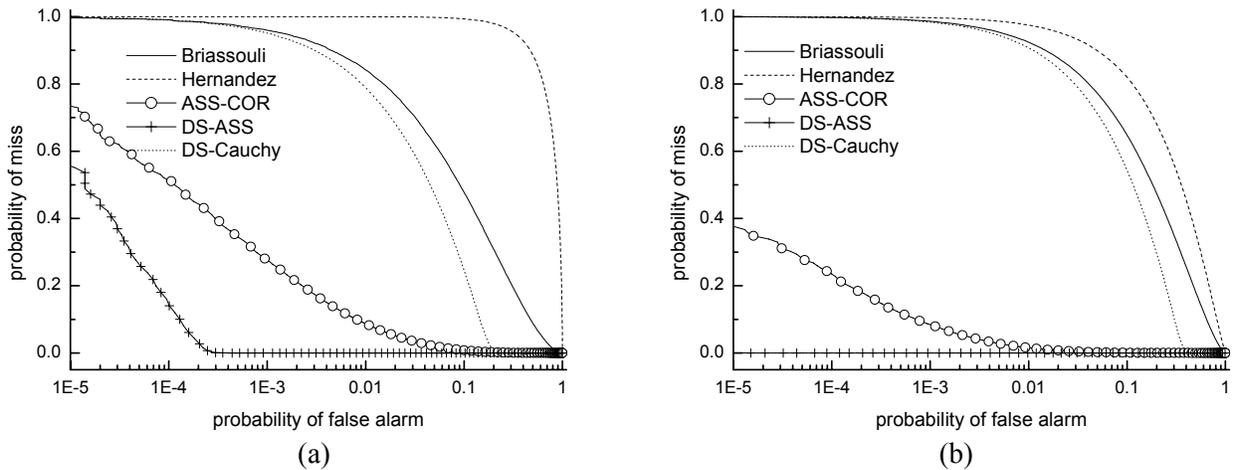

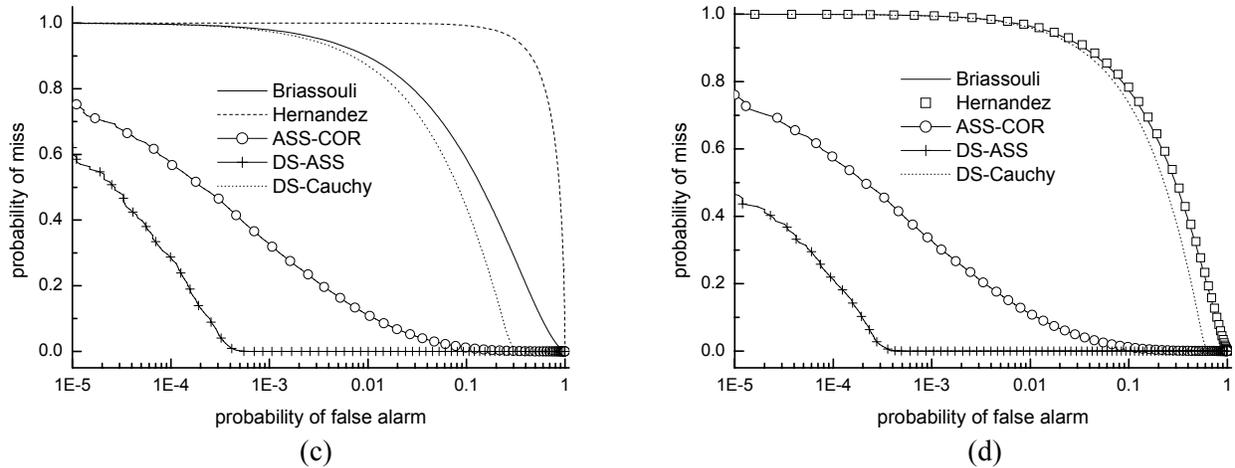

Fig. 4. Performance comparisons under no attack (with $a = 0.3$, $N = 2000$). (a) Lena. (b) Peppers. (c) Boat. (d) Baboon.

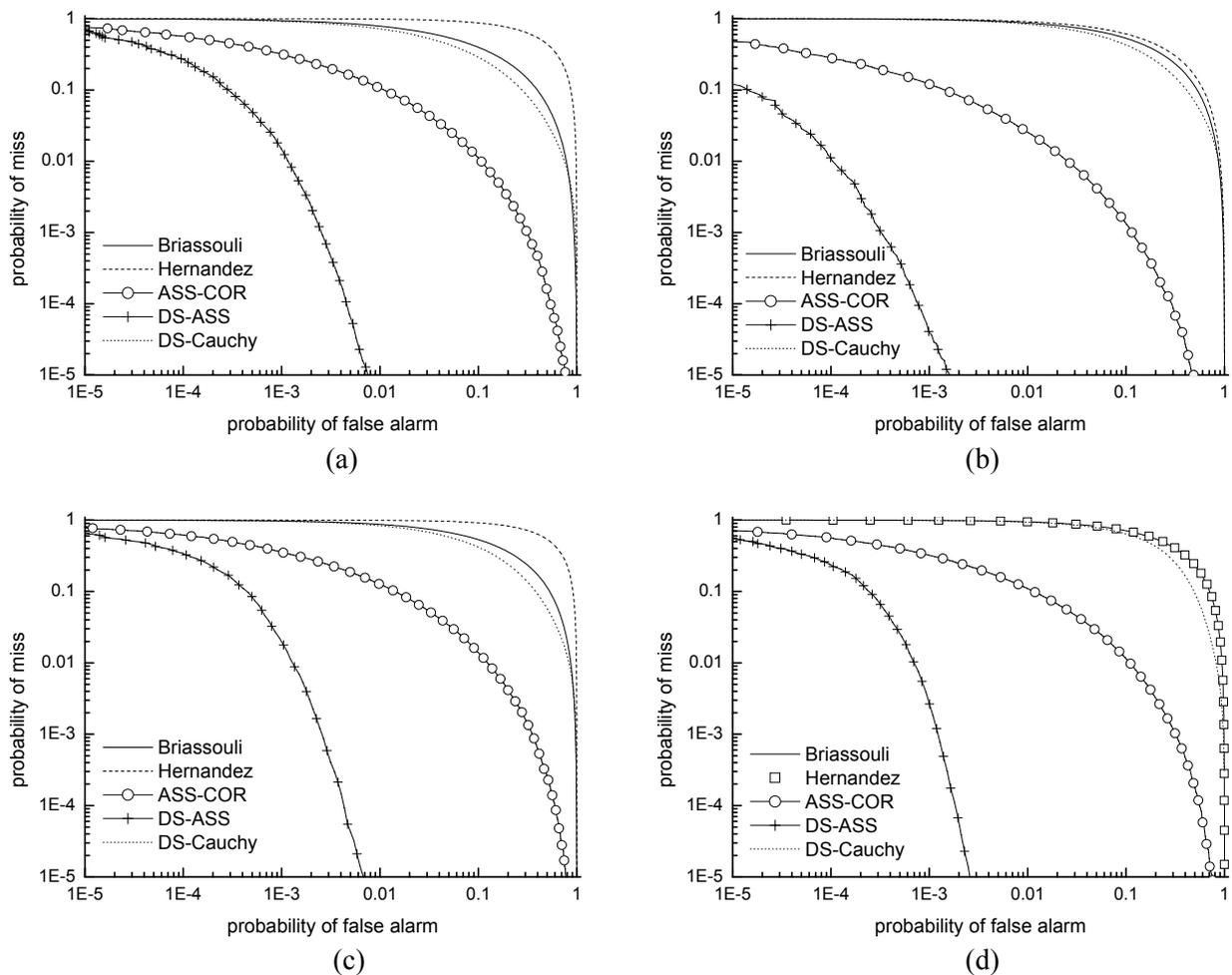

Fig. 5. Performance comparisons under zero-mean Gaussian noise attacks $\mathcal{N}(0, 25)$ (with $a = 0.3$, $N = 2000$). (a) Lena. (b) Peppers. (c) Boat. (d) Baboon.

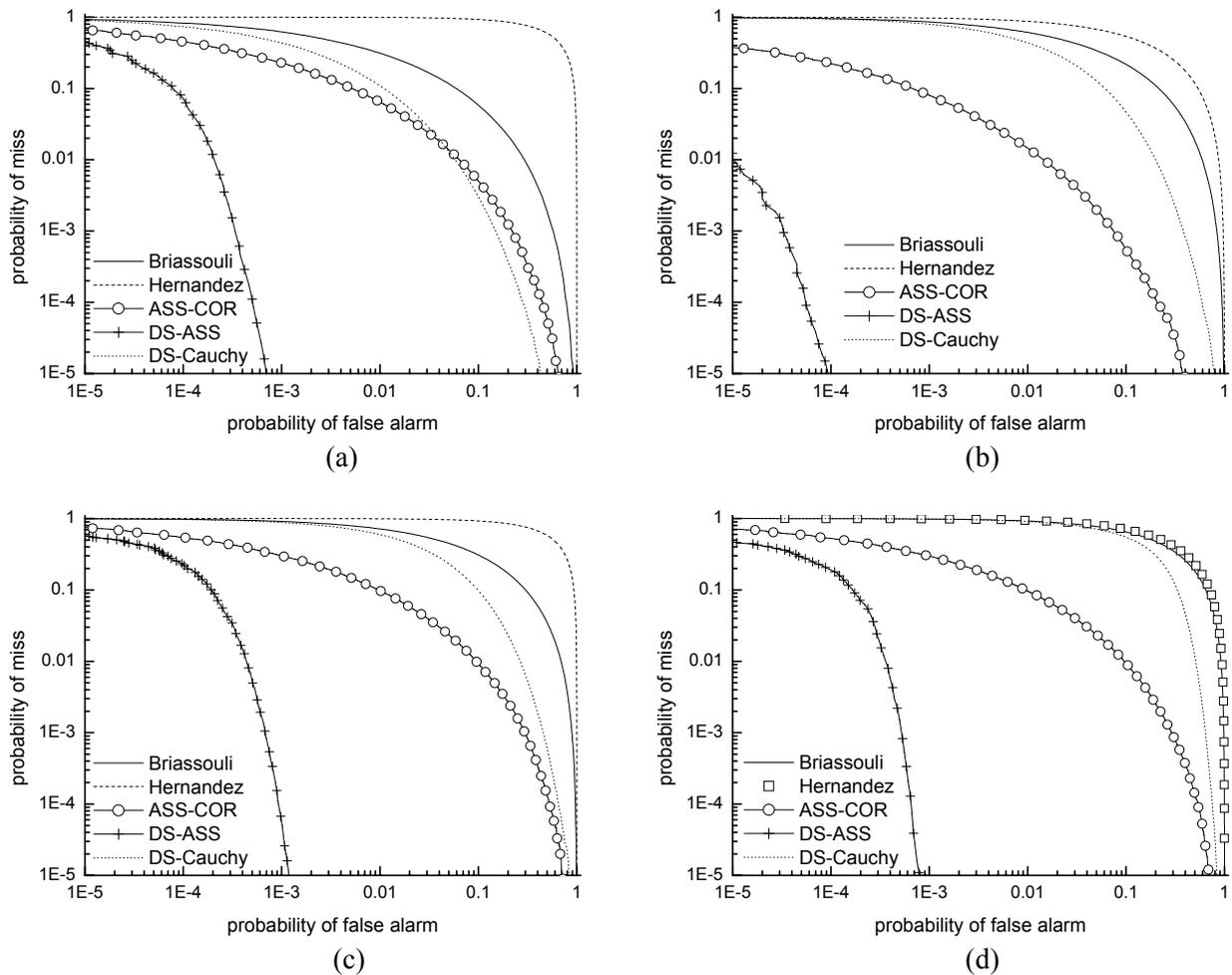

Fig. 6. Performance comparisons under JPEG (QF = 50) attacks (with $a = 0.3$, $N = 2000$). (a) Lena. (b) Peppers. (c) Boat. (d) Baboon.

In this subsection, the contrast masking is also included in the perceptual analysis. The comparison results are demonstrated in Fig. 4 to Fig. 6. Hernandez's and Briassouli's schemes are very poor in performance since the perceptual masks $m_i$ (with contrast masking) depend on host signals $x_i$ and thus the proposed detectors are not optimal. In fact, their performance is even much worse than that of the correlator. Second, as expected, DS-Cauchy still achieves a nicer performance than Briassouli's single-sided scheme. Third, most important of all, DS-ASS yields a much better performance than both ASS-COR and DS-Cauchy. This thus discourages the use of DS-Cauchy in real scenarios since it will leave a larger perceptual allowance for the attacker if the contrast masking is not implemented.

## VIII. Conclusions

In this work, we further demonstrated the advantage of double-sided schemes over its single-sided counterparts in a scenario where the Watson's perceptual model was implemented to improve the fidelity of the watermarked contents. Through performance comparisons, we find that DS-ASS is still the most appealing scheme. Even if the contrast masking is not implemented, DS-ASS is not necessary inferior to DS-Cauchy in performance. However, if the contrast masking is implemented, it offers a dramatic performance advantage over DS-Cauchy. Moreover, DS-ASS is applicable to almost all kinds of host data whatever their probabilities of distributions are. Thus, it is also applicable to audio and video watermarking.

## Acknowledgements

The authors would like to thank the three anonymous reviewers and Prof. Upamanyu Madhow for their insightful comments and suggestions. Their suggestions have been incorporated into this work.